# Interaction energy between two charged spheres surrounded by electrolyte. The smaller sphere located inside the larger sphere

István P. Sugár

Department of Neurology, Icahn School of Medicine at Mount Sinai, New York, NY 10029

## Abstract

By using the recently generalized version of Newton's Shell Theorem [3] analytical equations are derived to calculate the electric interaction energy between two charged spheres, a small one is located within a larger one, and each sphere is surrounded outside and inside by electrolyte.  This electric interaction energy is calculated as a function of the electrolyte's ion concentration, temperature, distance between the centers and size of the spheres. At the same distance between the centers of the spheres the absolute value of the interaction energy decreases with increasing electrolyte ion concentration and increases with increasing temperature. At zero electrolyte ion concentration the derived analytical equation transforms into the result of the Shell Theorem. Finally, the analytical equation is generalized to calculate the total electric interaction energy between a large charged sphere and N small charged spheres (located within the large sphere) where the spheres are surrounded by electrolyte.

## Introduction

Anionic phospholipids, such as phosphatidic acid (PA), phosphatidylserine (PS), phosphatidylethanolamine (PE), and phosphatidylinositol (PI), are located primarily in the inner leaflet of the plasma membrane [1].  Eukaryotes usually have a single nucleus, but a few cell types, such as mammalian red blood cells, have no nuclei, and a few others including osteoclasts have many [2]. The nuclei are enveloped by double layer of lipid membranes [2] which may contain charged lipids too.

In this paper we model a cell with one or many nuclei by a large charged sphere containing one or many small charged spheres where each sphere is surrounded inside and outside by electrolyte. By using the recently generalized Shell Theorem [3] analytical equations are derived to calculate the electric interaction energy between a large charged sphere and small charged sphere(s) (located inside the large charged sphere).

## Model

In Figure 1 a large charged sphere of radius $R_2$ is shown containing a small charged sphere of radius $R_1$ and the charges are distributed homogeneously on the surface of the spheres.

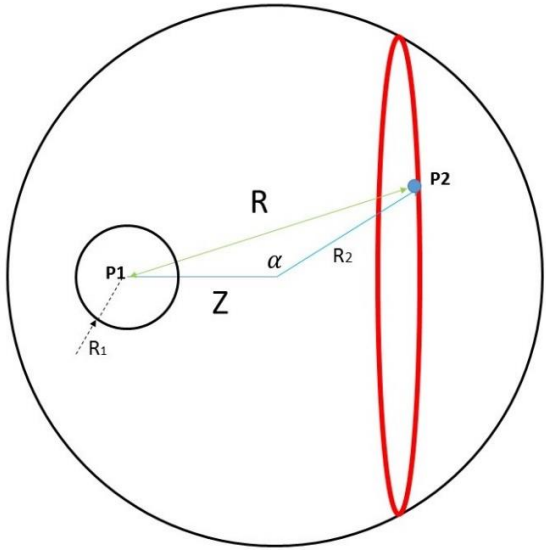

**Figure 1.** *Two charged spheres-the smaller sphere is located inside the larger sphere*
Small circle represents a charged sphere of radius $R_1$ and its total surface charge is $Q_1$. Large circle represents a charged sphere of radius $R_2$ and its total surface charge is $Q_2$. The distance between the centers of the two spheres is $Z$. The potential created by the small charged sphere is calculated at point P2 (located on the surface of the large sphere).
Red ring represent charges on the large charged sphere. Their distance from point P1 (the center of the small sphere) is R. $\alpha$ is the angle between vector $Z$ and a vector pointing from the center of the large sphere to any of the point (such as P2) of the red ring.

Based on the generalized Shell Theorem the potential created by the small charged sphere at point P2 is (Eq.9 in ref.3):

$$V(R) = \frac{k_e Q_1 \lambda_D}{\varepsilon_r R R_1} e^{-\frac{R}{\lambda_D}} \sinh\left(\frac{R_1}{\lambda_D}\right) \tag{1}$$

where $Q_1$ is the total charge of the small sphere, $\lambda_D$ is the Debye length in the electrolyte that is inside and around the charged spheres, $k_e$ is the Coulomb's constant and $\varepsilon_r$ is the relative static permittivity of the electrolyte.
The distance between point P1 and any of the point charges located on the red ring is:

$$R(\alpha, Z, R_2) = \sqrt{(R_2 \sin(\alpha))^2 + (Z - R_2 \cos(\alpha))^2} = \sqrt{R_2^2 + Z^2 - 2ZR_2 \cos(\alpha)} \tag{2}$$

The interaction energy between the small charged sphere and the charges of the red ring is:

$$E(\alpha)d\alpha = V(R)\rho_2 2R_2 \sin(\alpha) \pi R_2 \cdot d\alpha = \frac{k_e Q_1 \lambda_D}{\varepsilon_r R(\alpha,Z,R_2) R_1} e^{-\frac{R(\alpha,Z,R_2)}{\lambda_D}} \sinh\left(\frac{R_1}{\lambda_D}\right) \frac{Q_2}{2} \sin(\alpha) \cdot d\alpha \quad (3)$$

where $2R_2\sin(\alpha)\pi R_2 \cdot d\alpha$ is the surface area of the red ring, $\rho_2$ is the surface charge density on the large sphere and $Q_2 = 4R_2^2 \pi \rho_2$ is the total charge of the large sphere.

Finally, the interaction energy between the small and large sphere is:

$$E = \int_0^\pi E(\alpha)d\alpha = A \int_0^\pi \frac{\sin(\alpha)}{\sqrt{R_2^2+Z^2-2ZR_2\cos(\alpha)}} e^{-\sqrt{R_2^2+Z^2-2ZR_2\cos(\alpha)}/\lambda_D} d\alpha \quad (4)$$

where $A = \frac{k_e Q_1 \lambda_D}{\varepsilon_r R_1} \sinh\left(\frac{R_1}{\lambda_D}\right) \frac{Q_2}{2}$.

Let us make the following substitution in the integral: $u = \cos(\alpha)$.
Thus in Eq.4 $\sin(\alpha)\, d\alpha$ can be substituted by $-du$ and we get

$$E = A \int_{-1}^{1} \frac{1}{\sqrt{R_2^2+Z^2-2ZR_2 u}} e^{-\sqrt{R_2^2+Z^2-2ZR_2 u}/\lambda_D} du \quad (5)$$

Finally, let us make this substitution in Eq.5 : $w = -\sqrt{R_2^2 + Z^2 - 2ZR_2 u}/\lambda_D$ and thus

$$dw = \frac{ZR_2}{\lambda_D \sqrt{R_2^2+Z^2-2ZR_2 u}} du$$ and we get

$$E = \frac{A\lambda_D}{ZR_2} \int_{w(u=-1)}^{w(u=1)} e^w\, dw = \frac{A\lambda_D}{ZR_2} [e^w]_{w(u=-1)}^{w(u=1)} \quad (6)$$

where

$$w(u=-1) = -\frac{\sqrt{R_2^2+Z^2+2ZR_2}}{\lambda_D} = -\frac{(Z+R_2)}{\lambda_D}, \quad (7)$$

while in the case of $Z < R_2$

$$w(u=1) = -\frac{\sqrt{R_2^2+Z^2-2ZR_2}}{\lambda_D} = -\frac{\sqrt{(R_2-Z)^2}}{\lambda_D} = -\frac{(R_2-Z)}{\lambda_D}, \quad (8)$$

Thus when $Z < R_2 - R_1$, i.e. when the smaller sphere is inside the larger one:

$$E(Z) = \frac{A\lambda_D}{ZR_2} e^{-\frac{R_2}{\lambda_D}} \cdot 2 \cdot \sinh\left(\frac{Z}{\lambda_D}\right) = \frac{k_e Q_1 Q_2 \lambda_D^2}{\varepsilon_r R_1 R_2 Z} \sinh\left(\frac{R_1}{\lambda_D}\right) \sinh\left(\frac{Z}{\lambda_D}\right) e^{-\frac{R_2}{\lambda_D}} \quad (9)$$

## Results

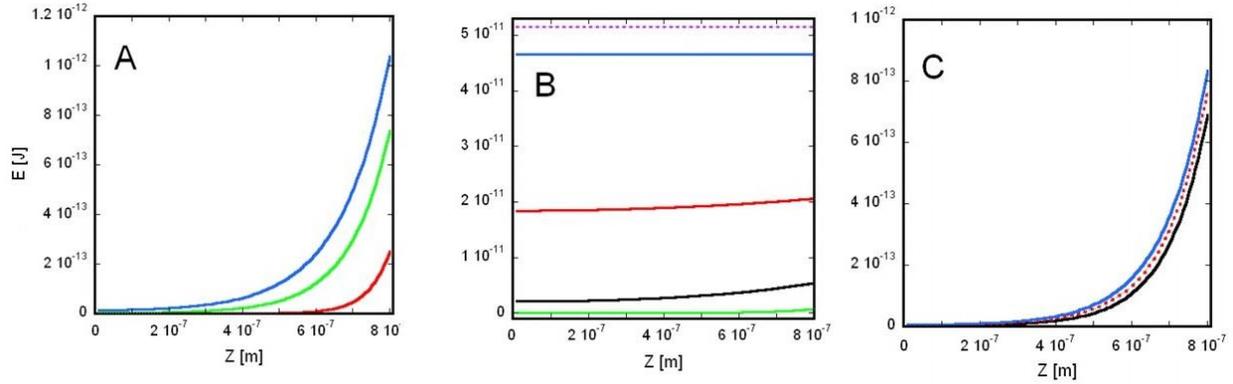

**Figure 2.** *Electric interaction energy of two charged spheres surrounded by electrolyte (dependence from electrolyte's ion concentration and temperature)*

The smaller sphere with radius $R_1 = 2 \cdot 10^{-7} m$ is located inside the larger sphere with radius $R_2 = 10^{-6} m$. The interaction energy between the two spheres, $E$ (Eq.9) is plotted against the distance between the centers of the two spheres, $Z$. A) The ion concentration of the electrolyte is: blue curve: $0.007 mol/m^3$ ($\lambda_D = 1.15 \cdot 10^{-7} m$); green curve: $0.01 mol/m^3$ ($\lambda_D = 9.62 \cdot 10^{-8} m$); red curve: $0.03 mol/m^3$ ($\lambda_D = 5.569 \cdot 10^{-8} m$); and the temperature in the case of each curve is $T = 300K$. B) The ion concentration of the electrolyte is: purple dotted curve: $0\ mol/m^3$ ($\lambda_D = \infty\ m$); blue curve: $0.000001 mol/m^3$ ($\lambda_D = 9.62 \cdot 10^{-6} m$); red curve: $0.0001 mol/m^3$ ($\lambda_D = 9.62 \cdot 10^{-7} m$); black curve: $0.001 mol/m^3$ ($\lambda_D = 3.04 \cdot 10^{-7} m$); green curve: $0.01 mol/m^3$ ($\lambda_D = 9.62 \cdot 10^{-8} m$) and the temperature in the case of each curve is $T = 300K$. C) The system's temperature is: blue curve: $340K$; red dotted curve: $310K$; black curve: $280K$ and the electrolyte's ion concentration in the case of each curve is $0.01 mol/m^3$. The respective Debye lengths are calculated by means of Eq.A2 in ref.4. In the case of our calculations the surface charge density of each charged sphere is $\rho_s = -0.266\ C/m^2$ and thus the total charge of the spheres are: $Q_1 = -1.33706 \cdot 10^{-13} C$ and $Q_2 = -3.34265 \cdot 10^{-12} C$.

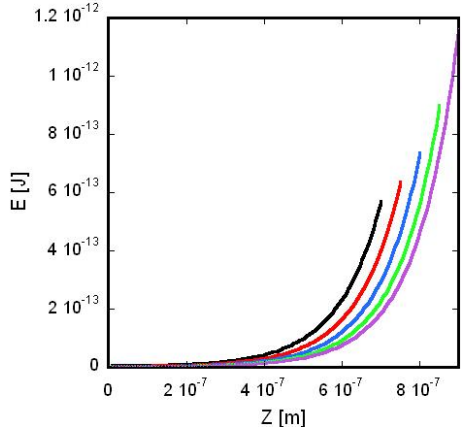

**Figure 3.** *Electric interaction energy of two charged spheres surrounded by electrolyte (dependence from radius)*

The interaction energy between the two spheres, $E$ (Eq.9) is plotted against the distance between the centers of the two spheres, $Z$. The total charge of the inner and outer sphere is $Q_1 = -1.33706 \cdot 10^{-13} C$ and $Q_2 = -3.34265 \cdot 10^{-12} C$, respectively. The radius of the large sphere (see Figure 1) is $R_2 = 10^{-6} m$, the electrolyte's ion concentration is $0.01 mol/m^3$, the temperature is $T = 300K$ and the Debye length (see Eq.A2 in ref.4) is $\lambda_D = 9.62 \cdot 10^{-8} m$. Purple curve: $R_1 = 10^{-7} m$; green curve: $R_1 = 1.5 \cdot 10^{-7} m$; blue curve: $R_1 = 2 \cdot 10^{-7} m$; red curve: $R_1 = 2.5 \cdot 10^{-7} m$; black curve: $R_1 = 3 \cdot 10^{-7} m$.

## Discussion

By using the recently generalized Shell Theorem, Eq.9 was derived to calculate the electric interaction energy between two charged spheres (where the smaller sphere is located inside the larger sphere) surrounded by electrolyte. In Eq.9 the interaction energy is given as the function of distance between the centers of the two spheres, $Z$. Because of the increased screening effect of the electrolyte's ions, at any given $Z$ distance between the spheres, the interaction energy decreases with increasing electrolyte ion concentration (Figure 2A,B). The primary reason of this decrease is that the last factor in Eq.9 ( $e^{-\frac{Z}{\lambda_D}}$ ) at a given $Z$ fast decreasing when the Debye length, $\lambda_D$ decreases (because of the increasing electrolyte ion concentration (see Eq.A2 in ref.4)). On the other hand $\lambda_D$ increases with increasing temperature (see Eq.A2 in ref.4) and thus at a given $Z$ factor $e^{-\frac{Z}{\lambda_D}}$ increases too causing the increase of the interaction energy between two charged spheres (see Figure 2C).

By increasing the radius $R_1$ the $E$ vs. $Z$ curves are shifting to the left (see Figure 3) because the highest value of $Z$ decreases, $Z_{max} = R_2 - R_1$. Also at $Z_{max}$ the electric interaction energy

$E(Z_{max})$ is getting smaller. This is the case because with increasing $R_1$ the distance between the charges of the spheres are increasing and thus the screening length increases too.

At infinite long Debye length (that is characteristic for vacuum) Eq.9 becomes:

$$E(Z) = \frac{k_e Q_1 Q_2}{\varepsilon_r Z} \left\{ \lim_{\lambda_D \to \infty} \frac{\lambda_D}{R_1} \left[ \frac{R_1}{\lambda_D} + \frac{1}{3!} \left(\frac{R_1}{\lambda_D}\right)^3 + \frac{1}{5!} \left(\frac{R_1}{\lambda_D}\right)^5 + \cdots \right] \right\} \left\{ \lim_{\lambda_D \to \infty} e^{-\frac{R_2}{\lambda_D}} \frac{\lambda_D}{R_2} \left[ \frac{Z}{\lambda_D} + \frac{1}{3!} \left(\frac{Z}{\lambda_D}\right)^3 + \frac{1}{5!} \left(\frac{Z}{\lambda_D}\right)^5 + \cdots \right] \right\} = \frac{k_e Q_1 Q_2}{\varepsilon_r R_2} \quad (10)$$

This agrees with the result of the Shell Theorem, i.e. the potential inside the large charged sphere is constant and equal with $\frac{k_e Q_2}{\varepsilon_r R_2}$ (see also the purple dotted line in Figure 2B).

Finally, one can calculate the total electric interaction energy between N+1 charged spheres, where N small spheres are located inside a large sphere as follows:

$$E = \sum_{\substack{i=1 \\ i \neq j}}^{N} \sum_{j=1}^{N} \frac{k_e Q_i Q_j \lambda_D^2}{\varepsilon_r R_i R_j Z_{ij}} \sinh\left(\frac{R_i}{\lambda_D}\right) \sinh\left(\frac{R_j}{\lambda_D}\right) e^{-\frac{Z_{ij}}{\lambda_D}} + \sum_{i=1}^{N} \frac{k_e Q_i Q_g \lambda_D^2}{\varepsilon_r R_i R_g Z_{ig}} \sinh\left(\frac{R_i}{\lambda_D}\right) \sinh\left(\frac{Z_{ig}}{\lambda_D}\right) e^{-\frac{R_g}{\lambda_D}} \quad (11)$$

where $Q_i$, $R_i$, is the total charge and radius of the $i$-th small sphere, respectively and $Z_{ij}$ (where $R_i + R_j < Z_{ij}$) is the distance between the centers of the $i$-th and $j$-th small sphere. The total charge and radius of the large sphere is $Q_g$ and $R_g$, respectively, while the distance between centers of the $i$-th small sphere and the large sphere is $Z_{ig}$ (where $0 < Z_{ig} < R_g - R_i$). The first term in Eq.11 is similar to Eq.12 in ref.4 calculating the total interaction energy between the N small spheres. The second term in Eq.11 calculates the interaction energy between the large sphere and each small sphere (i.e. sum of Eq.9 for the N small spheres).

## Conclusions

By using the recently generalized version of Newton's Shell Theorem [3] analytical equations are derived to calculate the electric interaction energy between two charged spheres, a small one is located within a larger one, and each sphere is surrounded outside and inside by electrolyte. This electric interaction energy is calculated as a function of the electrolyte's ion concentration, temperature, distance between the centers and size of the spheres. At the same distance between the spheres the absolute value of the interaction energy decreases with increasing electrolyte ion concentration and increases with increasing temperature. At zero electrolyte ion concentration the derived analytical equation transforms into the result of the Shell Theorem. Finally, the analytical equation is generalized to calculate the total electric interaction energy between a large charged sphere and N small charged spheres (located within the large sphere) where the spheres are surrounded by electrolyte.

## Acknowledgement

The author is very thankful for Chinmoy Kumar Ghose.

## References

1. Ma Y, Poole K, Goyette J, Gaus K (2017) Introducing Membrane Charge and Membrane Potential to T Cell Signaling, *Front. Immunol.* doi.org/10.3389/fimmu.2017.01513

2. Larijani B, Poccia DL (2009) Nuclear Envelope Formation: Mind the Gaps, *Annurev. Biophys.* 38:107-124

3. Sugár IP (2020) A generalization of the Shell Theorem. Electric potential of charged spheres and charged vesicles surrounded by electrolyte, AIMS Biophysics 7: 76-89

4. Sugár IP (2021) Interaction energy between two separated charged spheres surrounded inside and outside by electrolyte, arXiv:2103.13959v1